\documentclass[twocolumn, showpacs,preprintnumbers,amsmath,amssymb, nofootinbib]{revtex4}
\usepackage{graphicx}% Include figure files
\usepackage{dcolumn}% Align table columns on decimal point
\usepackage{bm}% bold math

\def\al{\alpha}
\def\be{\beta}
\def\de{\delta}
\def\ga{\gamma}

\def\ep{\epsilon}

\def\la{\lambda}
\def\ze{\zeta}

\def\Ga{\Gamma}

\def\La{\Lambda}

 \def\calC{{\hbox{\cal C}}}

 \def\one{{{\mathbb I}}}

 \def\D{{{\mathbb D}}}
 \def\C{{{\mathbb C}}}
 
 \def\E{{{\mathbb E}}}

 \def\F{{{\mathbb F}}}
 \def\G{{{\mathbb G}}}

 \def\K{{{\mathbb K}}}

\def\Aut{{\hbox{Aut}}}

\def\Spin{{\hbox{Spin}}}
\def\SO{{\hbox{SO}}}

\def\di{{\hbox{d}}}

\def\ip{\hbox to4pt{\leaders\hrule height0.3pt\hfill}\vbox to8pt{\leaders\vrule width0.3pt\vfill}\kern 2pt}
\def\QDE{\hfill\hbox{\ }\vrule height4pt width4pt depth0pt} 
\def\del{\partial}
\def\na{\nabla}

\def\arr{\rightarrow}

\def\then{\Rightarrow}

\def\ffrac[#1/#2]{\hbox{$\frac{#1}{#2}$}}
\def\Frac#1#2{\frac{#1}{#2}}
\def\({\left(}
\def\){\right)}
\def\[{\left[}
\def\]{\right]}
\def\^#1{{}^{#1}_{\>\cdot}}
\def\_#1{{}_{#1}^{\>\cdot}}
\def\<{\kern -1pt}

\begin{document}

\newpage

\title{Symmetry operators for Dirac's equation on two-dimensional  spin manifolds
\footnote{This paper is published despite the effects of the Italian law 133/08 ({\tt http://groups.google.it/group/scienceaction}). 
This law drastically reduces public founds to public Italian universities, which is particularly dangerous for scientific free research, 
and it will prevent young researchers from getting a position, either temporary or tenured, in Italy.
 One of the authors (L.F.) is protesting against this law to obtain its cancellation.\\}}
% Force line breaks with \\

\author{Lorenzo Fatibene}
 \email{lorenzo.fatibene@unito.it}
  \affiliation{Dipartimento di Matematica, Universit\`a di Torino, Italy\\
 INFN - Iniz.~Spec.~Na12}

\author{Raymond G.\ McLenaghan}
 \email{rgmclena@uwaterloo.ca}
 \affiliation{Department of Applied Mathematics, University  of Waterloo, Waterloo, Ontario, N2L~3G1, Canada}
 
\author{Giovanni Rastelli}
 \email{giorast.giorast@alice.it}
 \affiliation{Dipartimento di Matematica, Universit\`a di Torino, Italy}
 
\author{ Shane N.\ Smith}
 \email{shanesmith@yahoo.com}
% \affiliation{address unknown}
 

\begin{abstract}
%We obtain and classify the second order symmetry operators for the Dirac equation on a general
%two-dimensional spin manifold.
It is shown that the second order symmetry operators for the Dirac equation on a general
two-dimensional spin manifold may be expressed in terms of Killing vectors and valence 
two Killing tensors.  The role of these operators in the theory of separation of
variables for the Dirac equation is studied. 
%We shall study symmetry operators and classify all symmetry operators of Dirac equation on a generic
%2D spin surface. No assumption on the surface metric structure.
\end{abstract}


\maketitle

\def\FieldEqsEquivalenceAPP{A}
\def\FrameAPP{B}
\def \ProjectionIdentitiesAPP{C}

\section{Introduction}
Exact solutions to Dirac's relativistic wave equation by means of the method of separation of variables have been studied since the equation was postulated in 1928.  Indeed, the solution for the hydrogen atom may be obtained by this method.  While there is a well developed theory of separation of variables for the Hamilton-Jacobi equation, and the 
Shr\"{o}dinger equation based on the existence of valence two characteristic Killing tensors which define 
respectively quadratic first integrals and second order symmetry operators for these equations (see \cite{MILLER1,KALNINS,B97,FFFG})
%the second order linear partial differential equations of mathematical physics (see\cite{MILLER1,KALNINS}), 
an analogous theory for the Dirac equation is still in its early stages.  The complications arise from the fact that one is dealing with a system of first order partial differential equations whose derivation from the invariant Dirac equation depend not only on the choice of coordinate system but also on the choice of an orthonormal moving frame and  representation for the Dirac matrices with respect to which the components of the unknown spinor are defined.  Further complications arise if the background space-time is assumed to be curved.  Much of the progress in the theory  has been stimulated by developments in Einstein's general theory of relativity where one studies
%wishes to study  
first quantized relativistic electrons on curved background space-times of physical interest such as the Schwarzschild and Kerr black hole space-times.  This work required the preliminary development of a theory of spinors  on  general pseudo-Riemmanian manifolds (see \cite{FF}, \cite{CFF}, \cite{Jad}, \cite{Spi}).  The solution of the Dirac equation in the Reissner-Nordstrom solution was apparently first obtained by Brill and Wheeler in 1957 \cite{BW} who separated the equations for the spinor components in standard orthogonal Schwarzschild coordinates with respect to a moving frame adapted to the coordinate curves.  A comparable separable solution in the Kerr solution was found by Chandrasekhar in 1976 \cite{CHAN} by use of an ingenious separation ansatz involving Boyer-Lindquist coordinates and the Kinnersley tetrad.  The separability property was characterized invariantly by Carter and 
McLenaghan \cite{CM} in terms of a first order differential operator constructed from the valence two Yano-Killing tensor that exists in the Kerr solution, that commutes with the Dirac operator and that admits the separable solutions as eigenfunctions with the separation constant as eigenvalue.   Study of this example led Miller \cite{MILLER2} to propose the theory of a factorizable system for first order systems of Dirac type in the context of which the separabilty property may be characterized by the existence of a certain system of commuting first order symmetry operators.   While this theory includes the Dirac equation on the Kerr solution and its generalizations \cite{FK} it is apparently not complete since as is shown by Fels and Kamran \cite{FK} there exist systems of the Dirac type whose separablity is characterized by second order symmetry operators.  The work begun by these authors has been continued by Smith \cite{Sm},
Fatibene, McLenaghan and Smith \cite{FMS} and McLenaghan and Rastelli \cite{MR} who studied the problem in the simplest
possible setting namely on two-dimensional Riemanian spin manifolds.  
The motivation for working in
the lowest permitted dimension is that it is possible to examine in detail the different possible scenarios that arise from the separation ansatz and the imposition of the separation paradigm that the separation be characterized by a symmetry operator admitting the separable solutions as eigenfunctions.    
The insight obtained from this approach may help suggest an approach to take for the construction of a general separability theory for Dirac type equations.  Indeed in \cite{MR} systems of two first order linear partial equations of Dirac type which admit multiplicative separation of variables in some arbitrary coordinate system and whose separation constants are associated with commuting diffrential operators are exhaustively characterized.  The requirement
that the original system arises from the Dirac equation on some two-dimensional Riemannian spin manifold allows the local
characterization of the orthonormal frames and metrics admitting separation of variables for the equation and the determination of the symmetry operators associated to the separation.

The purpose of the present paper is to take this research in a different but closely related direction.  Following
earlier work of McLenaghan and Spindel \cite{MSp} and Kamran and McLenaghan \cite{KML} where the first order
symmetry operators of the Dirac equation where computed on four-dimensional Lorenzian spin manifolds and McLenaghan,
Smith and Walker \cite{MSW} where the second-order operators were determined in terms of a two-component spinor formalism, 
we obtain the most general second-order linear differential operator which commutes with the Dirac operator on a general 
two-dimensional Riemannian spin manifold and show that it is characterized in terms of Killing vectors and valence two Killing 
tensors defined on the background
manifold.  The derivation is manifestly covariant; we work in a general orthonormal frame and do not chose a particular
set of Dirac matrices.  The paper is organized as follows: In Section 2 we derive the conditions that must be satisfied by the
coefficients of a second order linear operator which commutes with the Dirac operator.  In Section 3 we expand each zero-order
coefficient operator in terms of a suitable basis of the Clifford algebra $\calC(2)$ and solve sequentially
the three defining conditions obtained in Section 2.  By this procedure we show that the general second order
symmetry operator is characterized by a valence two Killing tensor field, two Killing vector fields and two scalar fields defined on the
background spin-manifold.  These fields are related by a certain differential condition the integrability condition of which
is solved on the Liouville surface which is the most general surface which admits a general valence two Killing tensor.
In Section 4 we give theorems characterizing non-trivial first order symmetry operators in terms of Killing vectors 
and non-trivial second order symmetry operators in terms of characteristic valence two Killing tensors.  We also give
an explicit example of an non-trivial, irreducible Killing tensor which is defined on the asymmetrical ellipsoid
surface.  In Section 5 we establish a link between the second-order symmetry operators studied in the previous sections and the naive
separation of variables considered in \cite{MR} by exploiting the fact that the existence of a non-trivial second order symmetry operator
implies in general that the Liouville metric has at least one ignorable coordinate.  Section 6 contains the conclusion.
The notation and conventions of this paper  are consistent with \cite{MR}.
 
\section{Two-dimensional spin manifolds}

We fix Euclidean signature by setting $\eta=(2,0)$ and denote by $\eta_{ab}=\de_{ab}$ the corresponding  canonical symmetric tensor.
A representation of the Clifford algebra $\calC(2)$ is induced by a set of Dirac matrices $\ga_a$ such that
\begin{equation}
\ga_a \ga_b + \ga_b \ga_a =2 \eta_{ab} \one
\label{DiracIdentitiesEQ}
\end{equation}

We stress that we shall never fix a particular set of Dirac matrices, but we shall use only algebraic consequences of (\ref{DiracIdentitiesEQ}).
The even Clifford algebra is spanned by $\one$, $\ga_{a}$, and  $\ga:=\ga_1\ga_2$ and the most general element of the corresponding $\Spin(2)$ 
group is
\begin{equation}
S=a\one + b\ga
\quad\hbox{with } a^2+b^2=1
\end{equation}
One can define a covering map $\ell:\Spin(2)\arr \SO(2)$ by setting
\begin{equation}
\ell(S)=\left(\begin{matrix}
a^2-b^2		& 	-2 ab \\ 
 2 ab			& 	a^2-b^2\\ 
\end{matrix}\right)
\end{equation}

We stress that until now we are at a purely algebraic level.
Let $M$ be a connected, paracompact, $2$ dimensional spin manifold.
Let $P\arr M$ be a suitable $\Spin(2)$-principal bundle, such that it allows global maps $\La:P\arr L(M)$ of the spin bundle into the general frame bundle.
The local expression of such maps is given by spin frames $e_a^\mu$; see \cite{FF}. 
A spin frame induces a metric $g_{\mu\nu}= e^a_\mu\> \eta_{ab}\> e^b_\nu$ and a spin connection
\begin{equation}
\Ga^{ab}_\mu = e^a_\al \( \Ga^\al_{\be\mu} e^{b\be} + \partial_\mu e^{b\al}\)
\end{equation}
where $ \Ga^\al{}_{\be\mu}$ denotes the Levi-Civita connection of the induced metric $g_{\mu\nu}$.
We note that such a connection satisfies $\na_\mu e_a^\nu=0$.
We also remark that Latin indices are raised and lowered by the inner product $\eta_{ab}$ while Greek indices are raised and lowered by the induced metric 
$g_{\mu\nu}$.
For subsequent use we introduce the frame covariant derivative $\na_a:=e_a^\mu \na_\mu$. 
The Dirac equations then has the form
\begin{equation}
\D\psi= i \ga^a \na_a \psi -m\psi =0,
\label{DiracEquationEQ}
\end{equation}

where the covariant derivative of the spinor $\psi$ is defined as 
\begin{equation}
\na_\mu \psi= \di_\mu \psi +\ffrac[1/4] \ep_{ab} \Ga^{ab}_\mu \ga\> \psi
\end{equation}
and $\ep_{ab} $ denotes the permutation symbol. 

A spin transformation is an automorphism of the spin bundle $P$, i.e.~ locally
\begin{equation}
\begin{cases}
 x'= f(x) \\
 g'= S(x)\cdot g\\
\end{cases}
\end{equation}
They form a group denoted by $\Aut(P)$ which acts on spinors, frame and spin connection as
\begin{equation}
\begin{cases}
\psi'= S\cdot\psi \\
e'{}_a^\mu= J^\mu_\nu e_b^\nu \ell_a^b(S)\\
\Ga'{}^{ab}_\mu= \bar J_\mu^\nu \ell^a_c(S)\( \Ga^{cd}_\nu \ell^b_d(S) + \di_\nu \ell^c_d(S)\> \eta^{db}\)\\
\end{cases}
\end{equation}
leaving the Dirac equation (\ref{DiracEquationEQ}) invariant. 
Here $J^\mu_\nu$ is the Jacobian matrix of the spacetime transfomation $x'= f(x)$ onto which the spin transformation projects and $\bar J^\mu_\nu$ denotes the anti-Jacobian.

A {\it second order symmetry operator} for the Dirac equation is an operator of the form
\begin{equation}
\K= \E^{ab}\na_{ab} + \F^a\na_a +\G \one 
\label{29}\end{equation}
which commutes with the Dirac operator $\D$.
Here $\na_{ab} = \ffrac[1/2]\( \na_a \na_b + \na_b\na_a\)$ denotes the {\it symmetrized} second covariant derivative (expressed in the frame).
The coefficients $ \E^{ab},  \F^a, \G$ are matrix zero-order operators.

The symmetry equation $[\K, \D]=0$, can be easily expanded to the equivalent equations for the coefficients
\begin{widetext}
\begin{equation}
\begin{cases}
\E^{(ab} \ga^{c)} - \ga^{(c}\E^{ab)}   =0\\
\F^{(a} \ga^{b)} - \ga^{(b}\F^{a)}   = \ga^c\na_c \E^{ab}\\
\G \ga^{a} - \ga^{a}\G   = \ga^c\na_c \F^{a} 
-\ffrac[R/4]\( \E^{ab}\ga^c +\ga^c\E^{ab} \) \ep_{bc}\ga
+ \ffrac[R/6]\( \E^{bd}\ga^c +2 \ga^c \E^{bd}\) \ep^a{}_d \ep_{bc}
\\
\ga^a\na_a \G=
\ffrac[R/8]\( \F^{a}\ga^b +\ga^b\F^{a} \)\ga \ep_{ab}
+ \ffrac[1/12]\( 2\E^{ab}\ga^c +\ga^c\E^{ab} \)\ga \ep_{ac} \na_b R
\\
\end{cases}
\label{MainEQ}
\end{equation}
\end{widetext}
Here $R$ denotes the Ricci scalar of the induced metric. We observe that, in 2-dimensions, it encodes all the curvature information of the surface. In particular the frame components of the Riemann tensor may be written as
\begin{equation}
 R_{abcd} = \ffrac[R/2] \ep_{ab} \ep_{cd}
\end{equation}

\section{Second order symmetry operators}

In this Section we shall solve (\ref{MainEQ}). 
We begin by considering the first equation therein.
The coefficients $\E^{ab}$ are zero-order matrix operators which can be expanded in the basis of $\calC(2)$ defined above:
\begin{equation}
\E^{ab}= e^{ab}\one + e^{ab}_c \ga^c + \hat e^{ab}\ga,
\end{equation} 
where the coefficients $e^{ab}$, $e^{ab}_c$, $\hat e^{ab}$ are point functions in $M$.
Since the matrices $\one$, $ \ga^c$, $\ga$ are independent, the first equation in (\ref{MainEQ}) implies
\begin{equation}
\begin{cases}
e^{(ab}_d \ep^{c)d}=0\\
\hat e^{(ab} \ep^{c)d}=0\\ 
\end{cases}
\label{FirstEquationSystem}
\end{equation}
By multiplying the second equation of (\ref{FirstEquationSystem}) by $\ep_{de}$ one obtains directly
$\hat e^{(ab} \de^{c)}_d=0$, the trace of which gives $2\hat e^{ab} +\hat e^{ab} +\hat e^{ab}=0$.  It follows  immediately that
\begin{equation}
\hat e^{ab}=0
\end{equation}

The first  equation of (\ref{FirstEquationSystem}) may now be solved to obtain
\begin{equation}
e^{ab}_c \ga^c= 2 \al^{(a} \ga^{b)}
\end{equation}
Here $\al^a$ are the coefficients along the frame $e_a:=e_a^\mu \del_\mu$ of an arbitrary vector field $\al$.

Hence the most general $\E^{ab}$ which satisfies the first equation in (\ref{MainEQ}) is of the form
\begin{equation}
\E^{ab}= e^{ab}\one + 2 \al^{(a} \ga^{b)}
\end{equation}

We now consider the second equation of (\ref{MainEQ}).  By setting
\begin{equation}
\F^a = f^a\one + f^a_b \ga^b + \hat f^a\ga
\end{equation}
it may be expanded to give
\begin{equation}
\begin{cases}
\na^{(a} \al^{b)}=0\\
-2 f^{(a}_c \ep^{b)c}= -2 \na_c \al^{(a} \ep^{b)c}\\
-2 \hat f^{(a} \ep^{b)c}=  \na^d e^{ab}\\
\end{cases}
\label{SecondEquationSystem}
\end{equation}

The first of these equations are the Killing equations for the vector $\al^a$, which is hence a Killing vector or
possibly the zero vector.

The second equation in (\ref{SecondEquationSystem}) can be recast as $\(f^{(a}_c  - \na_c \al^{(a}\)\ep^{b)c}=  0$, from which
one can easily prove that the most general $\phi^a_c$ satisfying the equation  $ \phi^{(a}_c  \ep^{b)c}=  0$ is $\phi^a_c= A \de^a_c$,
where $A$ is an arbitrary function  on $M$. Hence we have
\begin{equation}
f^{a}_c  =  \na_c \al^{a} + A\de^a_c
\end{equation}

The third equation in  (\ref{MainEQ}) can easily be rewritten as
\begin{equation}
 \hat f^{a} \de^{b}_c +  \hat f^{b} \de^{a}_c=  \ep_{dc}\na^d e^{ab}
\label{SecondThirdEQ}\end{equation}
Taking the trace we obtain 
\begin{equation}
 \hat f^{a} = \ffrac[1/3] \ep_{bc} \na^b e^{ac}
\label{fDesitizedEQ}\end{equation}
which shows that $ \hat f^{a}$ is uniquely determined by $e^{ac}$.
However, (\ref{SecondThirdEQ}) contains six scalar equations, of which only one has been used.
The other five equations are exploited by back-substituting (\ref{fDesitizedEQ}) into (\ref{SecondThirdEQ}) to obtain
an equation for $e^{ac}$ alone, namely:
\begin{equation}
\ep_{ef} \na^e e^{af} \de^{b}_c +  \ep_{ef} \na^e e^{bf} \de^{a}_c=  3\ep_{dc}\na^d e^{ab}
\end{equation}
This is an integrability condition for $e^{ab}$ which one can show is equivalent to
\begin{equation}
\na^{(d} e^{ab)}=0
\label{Killingeq}\end{equation}
This equation implies that $e^{ab}$ is a Killing tensor which we shall from here on denote by $K^{ab}$. We note that (\ref{Killingeq}) 
is necessary condition for  (\ref{SecondEquationSystem}) to be true.

To summarize the most general solution of the first and second equations in ($\ref{MainEQ}$) is given by
\begin{equation}
\begin{cases}
\E^{ab}= K^{ab}\one + 2 \al^{(a} \ga^{b)}\\
\F^a= f^a\one+ (\ga^c\na_c \al^{a} + A\ga^a) + \ffrac[1/3] \ep_{bc} \na^b K^{ac}\ga \\
\end{cases}
\end{equation}
where
\begin{equation}
\begin{cases}
 \na^{(d} K^{ab)}=0\\ 
\na^{(a} \al^{b)}=0\\ 
\end{cases}
\end{equation}
We observe that there is yet no condition on the vector field $f^a$.

We now consider the third equation of ($\ref{MainEQ}$). Writing
\begin{equation}
\G= g\one + g_a \ga^a + \hat g\ga
\end{equation}
we obtain the equations
\begin{widetext}
\begin{equation}
\begin{cases}
\na_a A=0 \quad \then A\in \C\\
g_b=-\ffrac[R/4] \al_b\\
-2 \hat g \ep^{ab} =\na^b f^a 
+ \ffrac[1/3] \ep_{dc} \ep_e\^b \na^e\na^d K^{ac} 
+  \ffrac[R/2] K^{ab}
+  \ffrac[R/2] K^{cd} \ep_d\^a \ep_c\^b \\
\end{cases}
\end{equation}
\end{widetext}
The first equation implies that $A$ is constant, while the second uniquely determines $g_b$ in terms of $\al^a$ and the Ricci scalar. To completely use the information contained in the third, which consists of nine scalar equations, we consider separately its symmetric and antisymmetric parts.  They may
be written as
\begin{equation}
\begin{cases}
0 =\na^{(b} f^{a)} 
+ \ffrac[1/3] \ep_{dc} \ep_e\^{(b} \na^e\na^d K^{a)c} 
+  \ffrac[R/2] K^{ab}
+  \ffrac[R/2] K^{cd} \ep_d\^{(a} \ep_c\^{b)} \\
4 \hat g  =\ep_{ba}\na^b f^a 
- \ffrac[1/3] \ep_{dc}  \na_a\na^d K^{ac} 
\\
\end{cases}
\label{SPlusA}
\end{equation}
Since $K^{ac}$ is a Killing tensor one can prove $\ep_{dc}  \na_a\na^d K^{ac} \equiv 0$. 
Hence the second equation in (\ref{SPlusA}) implies that
\begin{equation}
 \hat g  = \ffrac[1/4]\ep_{ba}\na^b f^a
\end{equation}

Turning our attention to the first equation in (\ref{SPlusA}), we rewrite it as
\begin{equation}
\na^{(a} \( f^{b)} -\na_c K^{b)c}\)=0
\end{equation}
It follows that there exists a Killing vector $\ze^a$ such that $\ze^a=f^{a} -\na_c K^{ac}$ from which we obtain
\begin{equation}
f^{a} =\ze^a + \na_c K^{ac}
\end{equation}

Hence the general solution of the first three equations in (\ref{MainEQ}) is given by
\begin{equation}
\begin{cases}
\E^{ab}= K^{ab}\one + 2 \al^{(a} \ga^{b)}\\
\F^a=(\ze^a + \na_c K^{ac})\one+ (\ga^c\na_c \al^{a} + A\ga^a) + \ffrac[1/3] \ep_{bc} \na^b K^{ac}\ga \\
\G= g\one  -\ffrac[R/4] \al_b \ga^b+ \ffrac[1/4]\ep_{ba}\na^b \ze^a\ga
\end{cases}
\end{equation}
where 
\begin{equation}
\begin{cases}
 A\in \C \\ 
 \na^{(d} K^{ab)}=0\\ 
\na^{(a} \al^{b)}=0, \> \na^{(d} \ze^{b)}=0\\ 
\end{cases}
\end{equation}

It remains to consider the fourth equation in  (\ref{MainEQ}).
We would expect that this matrix equation would yield three equations, one each from $\one$, $\ga^a$, and $\ga$. 
However, in this case two of them are identically satisfied and the only additional condition is
\begin{equation}
\na_a g= -\ffrac[1/4] \na_b \(R K\_a{}^{b}\)
\quad\then
\partial_\mu g= -\ffrac[1/4] \na^\nu \(R K_{\mu\nu}\)
\label{IntegrabilityConditions}\end{equation}

This equation locally detemines $g$ if (and only if) the right hand side is a closed $1$-form.
We may investigate the circumstances under which this condition holds with the use of some further information about Killing tensors
on surfaces.
For example, if $M$ is flat the right hand side is trivially zero which implies that $g\in \C$ is constant. 
However, if we consider the non-flat case, the condition is in general not satisfied.
Its analysis requires knowledge regarding which spin manifolds $M$ admit non-trivial valence two Killing tensors.

Fortunately, this knowledge is available.  Indeed, one has the following result \cite{BMS}: \textit{a 2-dimensional
Riemannian space admits a non-trivial valence two Killing tensor if and only if it is a Liouville surface in
which case there exists a system of coordinates $(u,v)$ with respect to which the metric $g$ and Killing tensor
$K$ have the following forms:}
\begin{equation}
g=\(A(u)+B(v)\)\(\di u^2 + \di v^2\)
\end{equation}
\begin{equation}
K=\Frac{B(v)}{A(u)+B(v)} \del_u\otimes \del_u
-\Frac{A(v)}{A(u)+B(v)} \del_v\otimes \del_v=
K^{ab}\> e_a\otimes e_b
\label{323}\end{equation}
\textit{where $A$ and $B$ are arbitrary smooth functions.  Furthermore, the frame components of $K$are given by}
\begin{equation}
[K^{ab}]=diag(B(v),-A(u))
%|K^{ab}|=\left|\begin{matrix}
%B(v)&0\\ 
%0&-A(u)\\ 
%\end{matrix}\right|
\label{324}\end{equation}

Since in this case we know the form of the metric (and hence the scalar curvature $R$) and the Killing tensor, 
we may expand the integrability condition of equation (\ref{IntegrabilityConditions}) to obtain the following condition:
\begin{widetext}
\begin{equation}
(A+B)^2(A' B''' + A'''B') +6A'B'\((A')^2+ (B')^2\)-6A'B'(A+B)(A''+B'')=0
\label{IntCond}\end{equation}
\end{widetext}
where the primes denote differentiation with respect to the appropriate variable.

If at least one of the two functions $A$ and $B$ is constant, then
this integrability condition is trivially satisfied. 
In this case the Liouville surface $M$ is a surface of revolution, since it admits a Killing vector.  For example if $A$ is constant, then $u$ is cyclic with $\del_u$ the corresponding Killing vector.
In this case, equation (\ref{IntegrabilityConditions}) determines $g$ uniquely modulo a constant.

To complete the analysis of condition (\ref{IntCond}) we need to investigate whether there exist solutions with neither$A$ nor $B$ constant. Dividing by $A'B'$ when needed and differentiating we obtain the following separated equation:
\begin{equation}
\Frac{1}{B'}\( \Frac{1}{B'} \(\Frac{B'''}{B'}\)'\)' + \Frac{1}{A'}\( \Frac{1}{A'} \(\Frac{A'''}{A'}\)'\)'=0
\end{equation}
which we may easily integrate obtaining
\begin{equation}
\begin{cases}
(A')^2 = k A^4 + a_3 A^3 + a_2 A^2 +a_1 A+a_0\\ 
(B')^2 = -k B^4 + b_3 B^3 + b_2 B^2 +b_1 B+b_0\\ 
\end{cases}
\end{equation}
where $k$ is the separatiuon constant, and $a_i$ and $b_i$ are integration constants.

Substituting back for the second derivatives of (\ref{IntCond}) we obtain
\begin{equation}
3(a_3-b_3)(A+B)=2(a_2+b_2)
\end{equation}
It follows that there are two cases to be considered:
\begin{itemize}
\item Case I ($a_3\not= b_3$): then $A$ and $B$ must both be constant. This conclusion contradicts the original hypothesis that both $A'$ and $B'$ are 
non-vanishing.
\item Case II ($a_3 = b_3$): then $a_2=-b_2$. By substituting back into (\ref{IntCond}) we find that
$a_1=b_1$ and $a_0=-b_0$.
\end{itemize}

We conclude that in the \textit{generic case} at least one of the functions $A$ and $B$ must be constant.
The only other possiblities are the special cases where $A$ and $B$ are solutions of the following system
of differential equations:
\begin{equation}
\begin{cases}
(A')^2 = k A^4 + a_3 A^3 + a_2 A^2 +a_1 A+a_0\\ 
(B')^2 = -k B^4 + a_3 B^3 - a_2 B^2 +a_1 B -a_0\\ 
\end{cases}
\label{SpecialSolutions}
\end{equation}
These special cases include, for example, the solution $A(u)=a u^2$ and $B(v)= a v^2$ which corresponds
to the plane in parabolic coordinates.
They also account for all constant curvature surfaces and some additional Liouville surfaces. 
In fact, one can compute the Ricci scalar for special cases (\ref{SpecialSolutions})
obtaining
\begin{equation}
R=(A-B)k+ \ffrac[1/2]a_3
\end{equation}
We observe that by setting $k=0$ in (\ref{SpecialSolutions}) we obtain the surfaces of constant curvature
which includes the plane when $a_3=0$.
However, for $k\not=0$ we obtain more general surfaces.
Which special class of Liouville surfaces satisfies condition (\ref{SpecialSolutions})
remains to be investigated.

In the sequel, we shall consider the generic case in which either $A'=0$ or $B'=0$.
The details about special cases satisfying (\ref{SpecialSolutions}) will be considered in 
a forthcoming paper.

\section{First order operators and reducibility of second order operators}

First order operators are obtained by setting $e^{ab}=0$ and $\al^a=0$. The integrability condition is trivially satisfied and $g\in \C$.
Thus the most general first order operator has the form
\begin{equation}
\begin{cases}
\E^{ab}= 0\\ 
\F^a=\ze^a \one+ A\ga^a \\ 
\G= g\one + \ffrac[1/4]\ep_{ba}\na^b \ze^a\ga
\end{cases}
\qquad\hbox{where}
\begin{cases}
 A, g\in \C \\ 
\na^{(d} \ze^{b)}=0\\ 
\end{cases}
\end{equation}

Among first order operators we identify the {\it trivial} ones, namely the first order operators of the form
\begin{equation}
\al \D + \be \K_0
\end{equation}
for some zero order operator $\K_0 = g\one$.  The following theorem characterizes the trivial first order operators:

\medskip
\medskip
\noindent{\bf THEOREM: \sl
There is a one-to-one correspondence between non-trivial first-order symmetry operators and Killing vectors $\ze$ on $M$.
\QDE}
\medskip

The situation for second order symmetry operators is somewhat more complicated although it can be 
analyzed along similar lines.

We define the {\it trivial} second order symmetry operators as the ones of the form $\D \circ \K_1+\K'_1=\K_1\circ \D +\K'_1$, 
where $ \K_1$ and  $ \K'_1$denote any first order symmetry operators. Thus the most general non-trivial second order symmetry 
operator has the form
\begin{equation} 
\begin{cases}
\E^{ab}= K^{ab}\one \\ 
\F^a= \na_c K^{ac}\one + \ffrac[1/3] \ep_{bc} \na^b K^{ac}\ga \\ 
\G= g\one 
\end{cases}
\end{equation}
where
\begin{equation} 
\begin{cases}
 \na^{(d} K^{ab)}=0\\ 
 K^{ab}\not=\la \eta^{ab} \\ 
 \na_{[\mu}\( \na^\la \(RK_{\nu]\la}\)\)=0\\ 
\end{cases}
\end{equation}
and where $g$ a solution of equation (\ref{IntegrabilityConditions}).  This results is summarized in the following theorem: 

\medskip
\medskip
\noindent{\bf THEOREM: \sl
There is a one-to-one correspondence between non-trivial second order symmetry operators and
Killing tensors $K^{ab}$ on $M$ such that $K^{ab}\not=\la \eta^{ab}$
and $\na_{[\mu}\( \na^\la \(RK_{\nu]\la}\)\)=0$.
\QDE
}
\medskip

In particular, if we have a non-trivial second order symmetry operator on $M$, then
$M$ must admit a Killing tensor $K^{ab}$; hence $M$ is a Liouville surface (see \cite{BMS}).
According to the analysis done in Section 3, $M$ generically must admit
a Killing vector $\ze$ (corresponding to the cases $A'=0$ or $B'=0$) or, non-generically, it must be in the 
form (\ref{SpecialSolutions}).

\medskip

A {\it reducible} second order symmetry  operator is a linear combination of products of first order operators, namely  $\la^{ij}\>\K^1_i \circ \K_j{}^1$ for any first order symmetry operators $\K^1_j$.

If we consider the product of two first order operators with $\F^a_{(i)}=\ze^a_{(i)} \one+ A_{(i)}\ga^a$ ($i=1,2$), we obtain the second order operator
defined by
\begin{equation}
K^{ab}= \ze_1^{(a}\ze_2^{b)} + 2 A_1 A_2\eta^{ab}
\quad
\al^a= \ffrac[1/2]\( A_1\ze^a_2 + A_2\ze^a_1\)
\end{equation}
One can easily check that, since $\ze_i^a$ are Killing vectors, $K^{ab}$  is automatically a Killing tensor and that $\al^a$ is a Killing vector, since it is a linear combination of Killing vectors.

On particular surfaces (for example, surfaces of constant curvature) all Killing tensors are known to be produced by linear combinations of symmetrized products of Killing vectors.  Thus on such surfaces all non-trivial second order symmetry operators are reducible.
To have non-trivial irreducible second order symmetry operators one needs Killing tensors other than the ones which are symmetrized products of Killing vectors.

The conditions for the existence of irreducible Killing tensors can be shown to be met on a explicit example. Let $M_e$ an asymmetric ellipsoid endowed with the metric induced by that of the Euclidean Space. 
As shown by Jacobi, the Hamilton-Jacobi equation of the geodesics of $M_e$ admits orthogonal separable coordinates. 
Therefore, also a symmetric Killing 2-tensor $K_e$ associated with them exists (see for example \cite{B97}). 
However, no Killing vector can exist on $M_e$.  It follows that $K_e$ is irreducible. 

Indeed, let us consider
\begin{equation}
\ffrac[x^2/u_i^2-a^2]+ \ffrac[y^2/u_i^2-b^2]+ \ffrac[z^2/u_i^2-c^2]=1
\label{Ell}\end{equation}
with $0\leq a^2<u_1^2<b^2<u_2^2<c^2<u_3^2$ and $(x,y,z)$ are Cartesian coordinates. The equations (\ref{Ell}) for $i=1\ldots 3$ define three families of confocal quadrics parametrized by $u_i$; $u_1$ and $u_2$ associated with one and two-folded hyperboloids, $u_3$ with the ellipsoids. By defining $M_e$ as $u_3=h$ for some  constant $h$, the variables $u_1$ and $u_2$ determine orthogonal separable coordinates on  $M_e$. From (\ref{Ell}) it is possible to write
\begin{equation}
\begin{cases}
x^2=-\ffrac[(a^2-h^2)(a^2-u_1^2)(a^2-u_2^2)/(a^2-b^2)(a^2-c^2)]\\ 
y^2= \ffrac[(b^2-h^2)(b^2-u_1^2)(b^2-u_2^2)/(a^2-b^2)(b^2-c^2)]\\ 
z^2=-\ffrac[(c^2-h^2)(c^2-u_1^2)(c^2-u_2^2)/(a^2-c^2)(b^2-c^2)]\\ 
\end{cases}
\end{equation}
and to obtain
\begin{equation}
\begin{cases}
g_{11}=-\ffrac[u_1^2(u_1^2-h^2)(u_2^2-u_1^2)/(u_1^2-c^2)(b^2-u_1^2)(a^2-u_1^2)]\\ 
g_{22}= \ffrac[u_2^2(u_2^2-h^2)(u_2^2-u_1^2)/(u_2^2-c^2)(b^2-u_2^2)(a^2-u_2^2)]\\ 
\end{cases}
\label{metric}
\end{equation}
It is evident that a rescaling $u=u(u_1)$, $v=v(u_2)$ of the coordinates puts the metric in Liouville form
\begin{equation}
g_{uu}=g_{vv}=A(u)+B(v)
\end{equation}
with the two functions $A$ and $B$ both non-constant.
Then, the coordinates $(u,v)$ are separable and the associated Killing tensor is given by (\ref{323}).
Such a Killing tensor is of course irreducible since the metric (\ref{metric}) admits no Killing vector.
We remark that the Killing tensor given in this case by (\ref{323}) does not satisfy condition (\ref{SpecialSolutions});
This is consistent with the fact that Dirac equation does not separate in this coordinate web (see \cite{MR}).  
Whether this feature, namely that existence of Killing vectors, is a necessary condition for the separation of the Dirac equation
and for the existence of second order symmetry operators also in the special cases (\ref{SpecialSolutions}) in addition to the generic case, 
will be investigated in a forthcoming paper.

\section{Second order symmetry operators and separation of variables}

It is known (\cite{Sm}, \cite{FMS} and references therein) that the Dirac equation separates only in cooordinate systems separating the Hamilton-Jacobi equation for the corresponding metric tensor, and only if one at least of these coordinates is associated to a Killing vector. In 2-dimensions, this is the case for the coordinate system $(u,v)$ for which  the metric is in Liouville form, as seen in Section 3, where, say, $u$ is an ignorable coordinate.  The appropriate spin-frame in this case is given by 
%In $(u,v)$ we write the Dirac equation and the second order symmetry operator $K$ (\ref{29}) with Killing tensor given by (\ref{324}), vectors $\al$ %and $\ze$ set to zero and with spin frame 
\begin{equation}
(e^{\mu} _a)=\( 
\begin{matrix}
0 & \ffrac [1/\sqrt{A(u)+B(v)}]\\ 
- \ffrac[1/\sqrt{A(u)+B(v)}] & 0 
\end{matrix} \),
\end{equation}
where $A=0$ and $B=\beta(v) ^{-2}$.
Assuming that vectors $\al$ and $\ze$ are zero the Dirac operator $\D$ and second order symmetry operator $\K$ may be written in matrix form as
%in form of matrix differential operators. After setting $A(u)=0$, $B(v)=\be(v)^{-2}$, we have
\begin{equation}
\D=\be \[
\( \begin{matrix}
0 &-1 \\  
1 & 0\end{matrix} \) \del_u
+\( \begin{matrix} 
i & 0 \\  
0 & -i\end{matrix} \)\del_v \] 
+ \ffrac[i/2]\( 
\begin{matrix}
-\be' & 0 \\  
0 & \be'\end{matrix}
\),
\end{equation}
where primes denote differentiation and
\begin{equation}
\K=\( \begin{matrix} 
1 & 0 \\ 
0 & 1
\end{matrix}  
\)\del_{uu}^2.
\end{equation}
These expressions for $\D$ and $\K$ coincide with the separation scheme $D_5$ and the corresponding symmetry operator $L_5$ found in \cite{MR}\footnote{In \cite{MR}, the right-hand side of each equation expressing quantities $X$ and $Y$ as functions of $A_i$, $B_i$ etc.,  from Section 7 on,  must be divided by 2; this correction is necessary to compare with notations used here and it affects all expressions derived from the formulae of above.  In particular, due to additional typos, the right-hand term of the last equation in (7.3) should be multiplied by $i\; k /2$.}
 obtained by analyzing naive separation for the Dirac equation in two-dimensional Riemannian manifolds. It is an instance of non-factorizable separation \cite{FK}. 
By computing 
\begin{equation}
\D\psi-m\psi=0
\end{equation}
and
\begin{equation}
\K\psi-\mu \psi=0
\label{55}
\end{equation}
with the separation ansatz
\begin{equation}
\psi=\( \begin{matrix} 
a_1(u)b_1(v) \\  
a_2(u)b_2(v)
\end{matrix}
\)
\end{equation}
after dividing the two components of $\D\psi$ by $\be a_1 b_2$ and $\be a_2b_1$ respectively, we obtain the separated equations
\begin{equation}
\begin{cases}
-\ffrac[a'_2/a_1]=-i \ffrac[b'_1/b_2]+ \ffrac[i\be'/2\be] \ffrac[b_1/b_2]+ \ffrac[m /\beta] \ffrac[b_1/b_2]\\ 
\ffrac[a'_1/a_2]=i \ffrac[b'_2/b_1]-\ffrac[i\be'/2\be] \ffrac[b_2/b_1]+ \ffrac[ m/\beta] \ffrac[b_2/b_1]\\ 
\end{cases}
\end{equation}
 and, from (\ref{55}) we have
\begin{equation}
\begin{cases}
a''_1=-\mu a_1\\ 
a''_2=-\mu a_2\\ 
\end{cases}
\label{C}
\end{equation}
By introducing separation constants $\mu_1$ and $\mu_2$ such that $\mu_1\mu_2=\mu$, the first system gives for the $a_i$
\begin{equation}
\begin{cases}
a'_2=-\mu_1a_1\\ 
a'_1=\mu_2a_2\\ 
\end{cases}
\label{B}
\end{equation}
and for the $b_i$
\begin{equation}
\begin{cases}
-ib'_1+i \ffrac[\beta'/2\beta]b_1+ \ffrac[m/\beta] b_1=\mu_1b_2\\ 
ib'_2-i \ffrac[\beta'/2\beta] b_2+ \ffrac[m/\beta] b_2=\mu_2b_1\\ 
\end{cases}
\label{A}
\end{equation}
It is evident that $\K$ provides, via (\ref{C}), a decoupling relation for the equations (\ref{B}). On the other hand, (\ref{C}) can be obtained by applying twice equations (\ref{B}). The symmetry operators associated with separation are generated by this means in (\cite{MR}). The solutions for the $a_i$ are
\begin{equation}
\begin{cases}
a_1=c_1\sin(\sqrt{\mu} u)+c_2\cos(\sqrt{\mu} u)\\ 
a_2=\sqrt{\ffrac[\mu_1/\mu_2] }[-c_2\sin(\sqrt{\mu} u)+c_1\cos(\sqrt{\mu} u)]\\ 
\end{cases}
\end{equation}
where $c_1, c_2\in \C$.

In general, the relations (\ref{A}), cannot be associated with symmetry operators (they can, for example, if $\be$ is a constant, that is for Cartesian coordinates). However, again it is possible to use them to obtain decoupled second-order differential equations in the $b_i$ and then the solutions of(\ref{A}) itself, in the same way as for the $a_i$. The solution obtained in this way is consistent; indeed, the separation constant appearing here is again $\mu=\mu_1 \mu_2$ and it is the eigenvalue of the symmetry operator $\K$. Therefore, no additional symmetry operator is necessary for the separation. The general solution of (\ref{A}) can be easily computed in Cartesian coordinates, $\be(v)=1$, and we have  
\begin{equation}
\begin{cases}
b_1=d_1\sin(Mv)+d_2\cos(Mv)\\ 
b_2= \ffrac[1/\mu_1][(d_1+iMd_2)\sin(Mv)+(d_2-iMd_1)\cos(Mv)]\\ 
\end{cases}
\end{equation}
where $d_1, d_2\in \C$ and $M= \sqrt{m^2-\mu}$.
It is remarkable that, even if the $a_i$ and the $b_i$ depend on the $\mu_j$, the products $\psi_i=a_ib_i$ depend on $\mu$ only. 

By setting, for example,  $\be$ equal to $1$, $e^v$, $\sinh(v)$, $\cosh(v)$, $k-\cos(v)$, with $k>1$, respectively, the corresponding Riemannian manifolds are: the Euclidean plane in Cartesian and polar coordinates, the sphere, the pseudo-sphere (or hyperbolic plane), the torus with metric induced from that of the Euclidean Space.

For polar coordinates in the Euclidean plane and in the cases of sphere and pseudosphere, equations (\ref{A}) can be  (formally) integrated with the use of computer algebra (Maple), to obtain solutions respectively in terms of Bessel functions and, on  the sphere and pseudo-sphere, in terms of hypergeometric functions.

To summarize, we obtained various examples of different behaviours with respect to Dirac equation separability. 
The skew ellipsoid is an example in which one has an (irreducible) Killing tensor and separable coordinates for Hamilton-Jacobi equation
though this Killing tensor is not associated to a second order symmetry operator for the Dirac equation; this is consistent with the fact 
that Dirac does not separate in these coordinates.
In the case of parabolic coordinates one has a Killing tensor, separable coordinates for Hamilton-Jacobi equation and a second order symmetry operator for Dirac, although the Dirac equation does not separable in these coordinates.
Finally, we presented several examples for which one has a Killing tensor, separable coordinates for the Hamilton-Jacobi equation 
and second order symmetry operators for which the Dirac equation is found to separate in the same coordinates.

\section{Conclusion}

We compute the most general second order symmetry operator for the Dirac equation on a 2-dimensional
spin manilfold.  Particular choices of coodinates, spin frames or Dirac matrices play no role in the calculations 
which is manifestly covariant.
The case of 2-dimensional Lorentzian spin manifolds will be considered in a subsequent paper.
We hope to use the insight obtained in this paper and in \cite{MR} to give an invariant characterization
of separability for Dirac equation comparable to that for the Schr\"odinger equation described, for example,
in \cite{KALNINS}.
It is unclear whether it will be possible to extend the same approach to higher dimensions, although we hope to be able to find new theoretical tools to tackle the problem.
The expectation is to be able to study the orbit of the group of spin transformations, which acts canonically on symmetry operators and consequently on the underlying separation schemes. This information is necessary in view of a classification of {\it non-equivalent} separation schemes. 
In fact, the action of spin transformation group defines an equivalence relation among symmetry operators which may define non-trivial equivalence between different schemes.

\section*{Acknowledgements}

The authors wish to thank their reciprocal institutions, the Dipartimento di Matematica, Universit\`a di Torino
and the Department of Applied Mathematics, University of Waterloo for hospitality during which parts of this
paper were written.  The research was supported in part by a Discovery Grant
from the Natural Sciences and Engineering Reasearch Council of Canada and by a Senior Visiting Professorships of the Gruppo Nazionale di Fisica Matematica - GNFM-INdAM (RGM).

%We wish to thank INdAM-GNFM for supporting visiting professor grant.
%This work is partially supported by MIUR: PRIN 2005 on {\it Leggi di conservazione e termodinamica in meccanica dei continui e teorie di campo}.  
%We also acknowledge the contribution of INFN (Iniziativa Specifica NA12) and the local research founds of Dipartimento di Matematica of Torino University.


\begin{thebibliography}{99}

\bibitem{B97} S. Benenti, ``Intrinsic characterization of the variable
  separation in the Hamilton-Jacobi equation,'' J.\ Math.\ Phys.\ {\bf 38},
  6578--6602 (1997).
  
\bibitem{BW} D.R.Brill and J.A.Wheeler: Interaction of neutrinos and gravitational fields, Rev. Mod. Phys. 29, 465-479, 1957.
  
\bibitem{BMS} A.T. Bruce, R.G. McLenaghan and R.G. Smirnov, ``A geometrical approach to the problem of
 integrability of Hamiltonian systems by separation of variables'', J.\ Geom.\ Phys. {\bf 39},
 301-322, (2001).
 
\bibitem{CM} B.Carter and R.G.McLenaghan: Generalized total angular momentum for the Dirac operator in curved space-time, Phys. Rev. D 19, 1093-1097, 1979.

\bibitem{CFF} M.Cavaglia, L.Fatibene, M.Francaviglia, {\it Two-dimensional dilaton gravity coupled to massless},
Classical Quantum Gravity 15 (1998), no. 11, 3627Ð3643. 


\bibitem{CHAN} S.Chandrasekhar: The mathematical theory of black holes, Clarendon, Oxford, 1983, 531.

\bibitem{FF} L.Fatibene and M.Francaviglia: Natural and gauge natural formalism for classical field theories. A geometric perspective including spinors and gauge theories, Kluwer, Dordrecht, 2003.

\bibitem{Jad} L.Fatibene, M.Francaviglia, {\it Deformations of spin structures and gravity},
Acta Physica Polonica B, 29 (4): 915-928 APR 1998.

\bibitem{Spi} L. Fatibene, M. Ferraris, M Francaviglia, M. Godina,
{\it Gauge formalism for general relativity and fermionic matter},
Gen. Relativity Gravitation 30 (1998), no. 9, 1371Ð1389.

\bibitem{FFFG} L. Fatibene, M. Ferraris, M. Francaviglia and R.G. McLenaghan,
``Generalized symmetries in mechanics and field theories'', {\em J.
Math. Phys.} {\bf 43} (2002), 3147-3161.
 
\bibitem{FMS} L.Fatibene, R.G.McLenaghan, and S.Smith: Separation of variables for the Dirac equation on low dimensional spaces. In Advances in general relativity and cosmology, Pitagora, Bologna, 2003, 109-127.

\bibitem{FK} M.Fels and N.Kamran: Non-factorizable separable systems and higher-order symmetries of the Dirac operator. Proc. Roy. Soc. London A 428, 229-249, 1990.

\bibitem{KALNINS} E.G.Kalnins: Separation of variables for Riemannian spaces of constant curvature, Longman, Harlow, 1986.

\bibitem{KML}
N. Kamran and R.G. McLenaghan, ``Symmetry operators for neutrino ans Dirac fields on curved spacetime'',
  Phys. Rev. D {\bf 30}, 357-362 (1984).
  
\bibitem{MR}
R.G. McLenaghan and G. Rastelli,
{\it  Separation of variables for systems of first-order partial differential equations and the Dirac equation in two-dimensional manifolds},
 IMA volumes in mathematics and its applications. M.Eastwood and W. Miller Jr. eds. Vol. 144, 471-496, Springer 2008
 
\bibitem{MSW}
R.G. McLenaghan and S.N. Smith, and D.M. Walker,
Symmetry operators for spin-1/2 relativistic wave equations on curved space-time,
Proc. Roy. Soc. London A 456, 2629-2643, 2000.
 
\bibitem{MSp}
R.G. McLenaghan and Ph. Spindel,
{\it Quantum numbers for Dirac spinor field on a curved space-time},
Phys. Rev. D 20, 409-413, 1979.

\bibitem{MILLER1} W.Miller,Jr.: Symmetry and separation of variables, Addison-Wesley,Reading, 1977.

\bibitem{MILLER2} W.Miller,Jr.: Mechanism for variable separation in partial differential equations and their relationship to group theory. In Symmetries and Nonlinear Phenomena, World Scientific, Singapore, 1988, 188-221.

\bibitem{Sm} S.Smith: Symmetry operators and separation of variables for the Dirac equation on curved space-times. PhD thesis, University of Waterloo, 2002.



\end{thebibliography}
\end{document}